\documentclass[12pt]{iopart}
\usepackage{graphicx, epsfig}
\begin{document}

\title[EXAFS study of Ni$_{50}$Mn$_{35}$In$_{15}$]{Local atomic arrangement and martensitic transformation in Ni$_{50}$Mn$_{35}$In$_{15}$: An EXAFS Study }
\author{P A Bhobe, K R Priolkar\footnote[3]{Author to whom any correspondence should be addressed} and P R Sarode}
 \address{Department of Physics, Goa University, Goa, 403 206 India.}

\ead{krp@unigoa.ac.in}
\date{\today}

\begin{abstract}
Heusler alloys that undergo martensitic transformation in ferromagnetic state are of increasing scientific and technological interest. These alloys show large magnetic field induced strains upon martensitic phase change thus making it a potential candidate for magneto-mechanical actuation. The crystal structure of martensite is an important factor that affects both the magnetic anisotropy and mechanical properties of such materials. Moreover, the local chemical arrangement of constituent atoms is vital in determining the overall physical properties. Ni$_{50}$Mn$_{35}$In$_{15}$ is one such ferromagnetic shape memory alloy that displays exotic properties like large magnetoresistance at moderate field values. In this work, we present the extended x-ray absorption fine-structure measurements (EXAFS) on the bulk Ni$_{50}$Mn$_{35}$In$_{15}$ which reveal the local structural change that occurs upon phase transformation. The change in the bond lengths between different atomic species helps in understanding the type of hybridization which is an important factor in driving such Ni-Mn based systems towards martensitic transformation. 
\end{abstract}
\submitto{\JPD}
\maketitle

\section{Introduction}
Ni-Mn based  Heusler alloys are magnetic materials that exhibit structural transformation of the martensitic type with large magnetic field induced strains and can be exploited for tremendous technological applications \cite{ulla, tick}. These materials also offer an excellent opportunity to investigate the various aspects of magnetic and structural phase transformations in a single system \cite{sod14, ent39}. As a magnetic field controlled shape memory effect is realized in these materials upon martensitic transformation, they are commonly reffered to as ``ferromagnetic shape memory alloys'' or FSMA. Recently, martensitic transformations were observed in ferromagnetic alloys with composition Ni$_{50}$Mn$_{50-x}$Z$_{x}$ where Z = In, Sn, Sb. The phase diagram of these systems indicate that with concentrations upto $\sim$ 16 at\% Z-element, a transformation from the high temperature austenitic phase to a low symmetry martensitic phase can be achived with the lowering of temperature \cite{suto}. Especially, the Sn and In alloy series have been in focus \cite{acet-nat, acet-72, acet-73, koya88, koya89,  yu} and are known to exhibit exotic properties that promise various innovative applications. A giant inverse magnetocaloric effect has been reported in Ni-Mn-Sn alloys (18 J/kg K at 5 T) where the application of the magnetic field adiabatically cools the sample \cite{acet-nat}. Whereas large magnetoresistance at moderate field values has been observed in Ni$_{50}$Mn$_{35}$In$_{15}$ at temperatures very close to room temperature \cite{yu}. In particular, Ni$_{50}$Mn$_{35}$Sn$_{15}$ has a cubic L2$_1$ structure at room temperature and transforms martensitically at T$_M \approx $ 200 K while, Ni$_{50}$Mn$_{35}$In$_{15}$ crystallizes in a B2 structure at room temperature and transforms to a layered structure consisting of orthorhombic and monoclinic stacks below T$_M \approx $ 295 K \cite{suto}. The investigation of the magnetic properties show that the overall magnetic exchange in both the parent and product phases is ferromagnetic with a Curie temperature (T$_C$) of 319 K for  Ni$_{50}$Mn$_{35}$Sn$_{15}$ \cite{acet-72} and 304 K for  Ni$_{50}$Mn$_{35}$In$_{15}$ \cite{acet-73}. In addition, the magnetization studies show that the structural transformation in these alloys could be induced by both temperature and magnetic field \cite{acet-73, koya88} placing them in the class of FSMA. 

For the development of new FSMA and to classify the best alloys for practical applications, certain important properties of these alloys have been mapped with their composition. Atleast for the much studied prototype Ni-Mn-Ga alloys a correlation between structural transformation temperature (T$_M$), martensitic crystal structure and average number of valence electrons per atom (e/a) has been established \cite{lans}. For the near-stiochiometric Ni-Mn-Ga alloys, the compositional dependence of T$_M$ is determined and it is found to increase linearly with an increase in e/a value \cite{vasil}. The role of e/a in relation to martensitic transformation temperature in Ni-Mn-Ga alloys has been discussed at length by Entel \etal \cite{ent39}. Keeping in view the dependence of T$_M$ on e/a it is interesting to compare the two systems: Ni$_{50}$Mn$_{35}$Sn$_{15}$ that has an e/a value of 8.05 and Ni$_{50}$Mn$_{35}$In$_{15}$ with e/a = 7.9. 
Inspite of Ni$_{50}$Mn$_{35}$Sn$_{15}$ having a slightly higher e/a value, it undergoes a martensitic transition at much lower temperature as compared to Ni$_{50}$Mn$_{35}$In$_{15}$. At such a point, understanding the correlation between the structural aspect of martensitic phase and the transformation temperature, T$_M$, becomes important. With this aim, Ni$_{50}$Mn$_{35}$In$_{15}$ has been studied for its structural properties in the austenitic and martensitic phase using {\it extended x-ray absorption fine structure} (EXAFS) measurements at Mn and Ni K-edges and compared with our results on Ni$_{50}$Mn$_{35}$Sn$_{15}$ \cite{pab2}. 

\section{Experimental Details} 
Ni$_{50}$Mn$_{35}$In$_{15}$ was prepared by repeated melting of the appropriate quantities of the constituent elements of 4 N purity under argon atmosphere in an arc furnace. The sample-bead so obtained was annealed at 800 K for 48 h followed by quenching in ice water. Subsequent energy dispersive x-ray analysis (EDX) confirmed the composition of the sample to be close to nominal with Ni = 50.78, Mn = 34.69 and In = 15.24. Small pieces were cut from the sample-bead using a low speed diamond saw and the remainder was crushed into very fine powder using a mortor and pestle. The temperature dependence of magnetization was studied using a Vibrating Sample Magnetometer with low field value of 100 Oe and in the temperature range 50 to 330 K. The room temperature crystal structure was determined by X-ray powder diffraction (XRD) recorded on Rigaku D-MAX IIC diffractometer with Cu K$\alpha$ radiation. A local Extended X-ray Absorption Fine Structure (EXAFS) at Ni and Mn K-edge were recorded at room temperature and liquid nitrogen temperature, in the transmission mode, on the EXAFS-1 beamline at ELETTRA Synchrotron Source using Si(111) as monochromator. Data analysis was carried out using IFEFFIT \cite{new} in ATHENA and ARTEMIS programs \cite{brav}. Theoretical fitting standards were computed with ATOMS and FEFF6 programs \cite{rav, zab}.

\section{Results}
\begin{figure}[h]
\centering
\epsfig{file=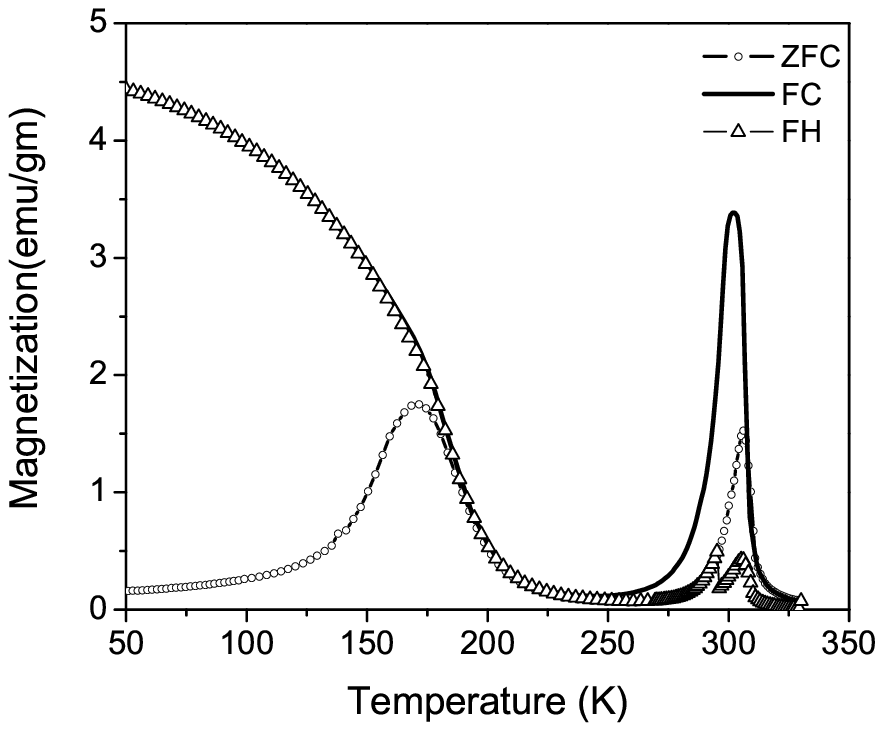, width=8cm, height=6cm}
\caption{\label{mag}Magnetization as a function of temperature measured in an applied field of 100 Oe between 50 to 330 K. The data was recorded while first warmup - ZFC, followed by cooling - FC and subsequent re-heating - FH.}
\end{figure}

M(T) measurements were carried out in a field of 100 Oe over the temperature range 50-330 K in the cooling/warming cycles as can be seen from the figure \ref{mag}. The first warmup followed by cooling and the subsequent heating data are indicated in the figure as ZFC (zero-field cooled), FC (field-cooled) and FH (field-heated) respectively. A steep rise in magnetization is observed at T$_C$ = 309 K in all the three data sets indicating a transition to magnetically ordered state. With the decrease in temperature a large seperation in the ZFC and FC curves indicative of competing magnetic interactions is observed. This feature is expected as the sample contains Mn atoms in excess to the stoichiometric composition giving rise to some antiferromagnetic interactions in addition to the strong ferromagnetic order. The ZFC magnetization increases slowly with the rise in temperature and forms a ``peak-like'' feature at $\sim$ 175 K. Such a feature is assigned to ferromagentic ordering of the martensitic phase \cite{acet-73}. A distinct splitting between the three curves is observed in the region 250K to 310K.This splitting can be attributed to the start and finish of martensitic transition. 

\begin{figure}[h]
\centering
\epsfig{file=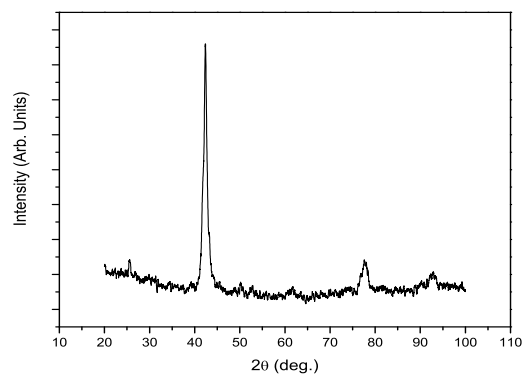, width=8cm, height=6cm}
\caption{\label{xrd}The x-ray powder diffraction pattern of Ni$_{50}$Mn$_{35}$In$_{15}$ recorded at room temperature. The absence of super-lattice reflections confirms the structure to be B2.}
\end{figure}

The XRD profile for Ni$_{50}$Mn$_{35}$In$_{15}$ is presented at figure \ref{xrd}. The absence of superlattice reflections corresponding to L2$_1$ order indicates that the structure is B2 with lattice parameter 3.0196$\pm$0.0006 \AA. Generally, in the cubic phase X$_2$YZ Heusler alloys are known to crystallize in the highly ordered L2$_1$ structure that consists of four face-centered cubic(fcc) sublattices. In the L2$_1$ unit cell the X atoms occupy the ($1\over4$, $1\over4$, $1\over4$) position, while the Y and Z atoms respectively occupy the (0, 0, 0) and (0, $1\over2$, 0) positions.  When the Y and Z atoms occupy the above mentioned sites absolutely at random causing a Y-Z disorder, the alloy is known to have a B2 structure. This disordered phase can be represented by a CsCl type structure with X atoms at the center of the cube and Y and Z atoms sharing the corners. In the case of Ni$_{50}$Mn$_{35}$In$_{15}$, the Ni atoms occupy the body centered position and the Mn, In atoms occupy the corners of the cube. Further, this alloy composition is non-stoichiometric with more Mn atoms than In which implies that Ni atoms have more number of Mn neighbours than In.

\begin{figure}[h]
\centering
\epsfig{file=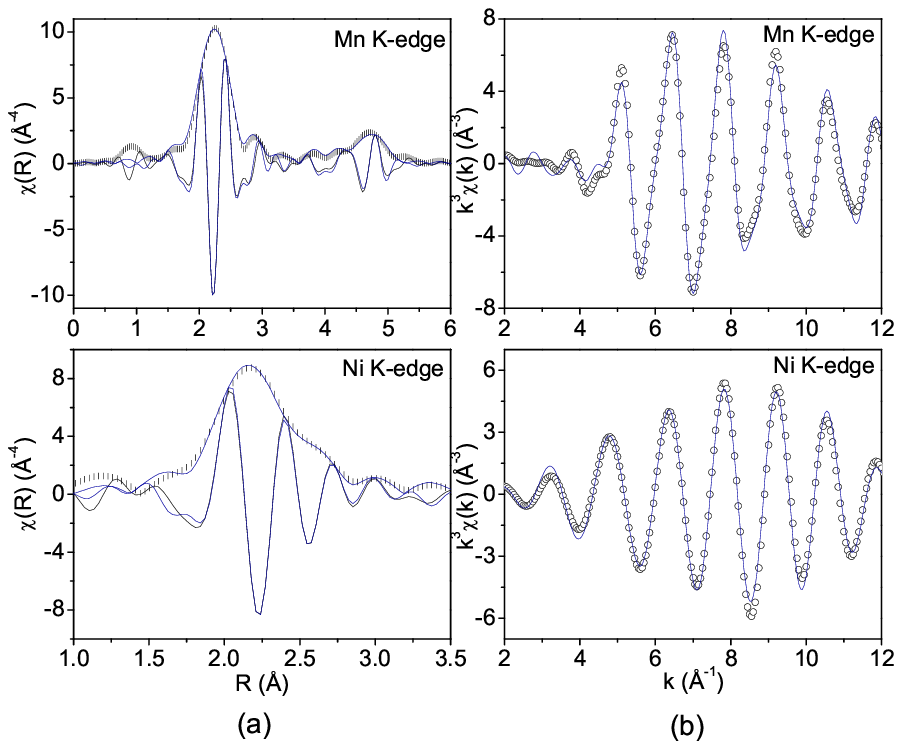, width=6cm, height=8cm}
\caption{\label{rt}(Color online) Magnitude and real component of Fourier Transform (FT) of EXAFS spectra in R space (left panel) and real component of FT in the back transformed k space (right panel) for Mn and Ni K-edge in Ni$_{50}$Mn$_{35}$In$_{15}$ obtained at room temperature. The fitting to the data are shown as coloured lines.}
\end{figure}

EXAFS were recorded at Mn and Ni K-edge in the austenitic and martensitic phases by carrying out the measurements at room temperature and liquid nitrogen temperature respectively. The room temperature EXAFS data was fitted with a cubic B2 crystal structural model. The coordination spheres involving Mn/In atoms as backscatters at the same bond distance were fitted as two separate correlations consisting entirely of  Mn and In atoms with coordination number fixed as per their composition ratio. A good fit was obtained for the values of the parameters mentioned in table \ref{tab:in-xf} and the fittings are shown in figure \ref{rt}. The parameters like bond length and the corresponding thermal mean square variation in the bond length for different correlations are obtained. The bond lengths calulated on the basis of the lattice parameters obtained from XRD profile are compared with values obtained from EXAFS analysis. Certain striking anomalies in the crystal structure are observed from the EXAFS analysis of both, the Mn and Ni edge data. From the analysis of the Mn edge data the second correlation representing the scattering between Mn (absorber) and Mn/In (backscatterer) atoms gives different values for Mn-In and Mn-Mn bonds. This difference in the bond distances is also reflected through the corresponding linear multiple scattering paths, directed along the body diagonal of the crystal lattice. Basically, the Mn-[In/Mn] bond represents the scattering between the Y-Z species of the X$_2$YZ alloy. For a stable cubic configuration and Mn as the absorbing atom such a correlation should have resulted in an equal Mn-In, Mn-Mn bond distance. Further, the third correlation at 4.27\AA~ comprising of yet another Mn-Mn bond has an exceptionally large $\sigma^2$ value. These discrepancies imply an unstable cubic structure with unequal movement of the constituent atoms from its crystallographic positions. As these anomalies are observed with respect to Mn atoms, it seems that in Ni$_{50}$Mn$_{35}$In$_{15}$, it is the Mn atom that moves considerably from its crystallographic position in comparison to the other constituents of the alloy. 

Room temperature Ni K-edge analysis also provides some interesting observations that hint towards the fact that Mn atoms are indeed moving. Table \ref{tab:in-xf} includes the parameters obtained from the analysis of the Ni K-edge data. Here, the fitting was carried out upto 3\AA~ in R space with three single scattering paths comprising of Mn, In and Ni as neighbours. When compared with the bond distance calculated from the XRD profile, it is seen that Ni-In and Ni-Ni bond distances match with the calulated values within the error bars while considerable deviation is observed for Ni-Mn bond distance. Further, this Ni-Mn bond distance differs from the value obtained from Mn EXAFS for the same (Mn-Ni) correlation. Ideally, the the Ni-Mn bond length as obtained from Ni K-edge and Mn K-edge data should have been equal as it involves the same pair of atoms. The difference in this correlation as observed from EXAFS analysis agrees with the argument that atoms move from their crystallographic positions over varied amplitudes of displacement with Mn atoms being displaced more than the other constituent atoms. Another striking feature in the Ni K-edge data is the difference in the Ni-Mn and Ni-In bond distance. If the crystal structure had to be perfectly cubic then Ni atoms being at the body-centered position, the Ni-In and Ni-Mn bonds should have been equidistant. The unequal Ni-Mn and Ni-In bond distance in  Ni$_{50}$Mn$_{35}$In$_{15}$ is related to the interplay between a static structural disorder and proximity to martensitic transformation temperature. It may be noted that the T$_M \sim$ 290 K for Ni$_{50}$Mn$_{35}$In$_{15}$ is close to the temperature of EXAFS measurement (295 K). The pre-transformational effects get very intense as T$_M$ is approached and any measurement in the narrow region of temperature around T$_M$ is known to be affected. The influence of such effects on the parameters obtained from EXAFS analysis has previously been observed75.50.Cc;  for Ni$_{2.1}$Mn$_{0.89}$Ga that has a T$_M \sim$ 285 K \cite{pab}. Hence the room temperature EXAFS of Ni$_{50}$Mn$_{35}$In$_{15}$ reflect the anomalies associated with martensitic transformation like unequal movement of constituent atoms and discrepancies in the bond distances for similar atomic correlations. 

\begin{figure}[h]
\centering
\epsfig{file=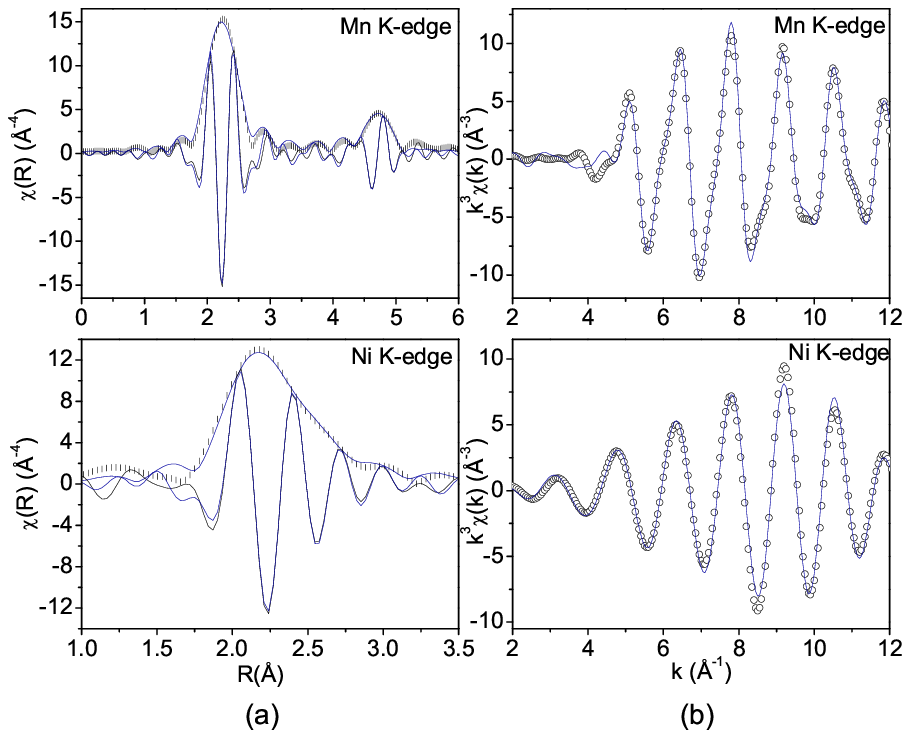, width=6cm, height=8cm}
\caption{\label{lt}(Color online) Magnitude and real component of FT of EXAFS spectra in R space (left panel) and real component of FT in the back transformed k space (right panel) for Mn and Ni K-edge in Ni$_{50}$Mn$_{35}$In$_{15}$ obtained at low temperature. The fitting to the data are shown as coloured lines.}
\end{figure}

The local structure of the martensitic phase of Ni$_{50}$Mn$_{35}$In$_{15}$ was fitted using correlations derived from the L1$_0$ structure. The bond distance and $\sigma^2$ values obtained from the low temperature EXAFS analysis are included in table \ref{tab:in-xf} and the fitted spectra are shown in figure \ref{lt}.  Martensitic transformation results in splitting of the bonds that were degenerate in the cubic cell due to lowering of crystal symmetry. Such splittings are evident from the Mn-Mn, Mn-Ni correlations in the Mn edge and Ni-Ni correlations in the Ni edge (see table \ref{tab:in-xf}). However, for the nearest neighbour interactions, the difference between the split bonds is very small and cannot be resolved by EXAFS. At such instances we have used single co75.50.Cc; rrelations in the fitings. For example in case of the first and second neighbour scatterings in both Mn and Ni EXAFS data. The first Mn-Ni bond distance as obtained from the Mn edge data remains unchanged from its room temperature value while from the Ni edge data the same bond distance decreases from 2.580\AA~ to 2.558\AA. As mentioned above, such differences in bond distances involving the same pair of atoms can be attributed to the  movement of atoms from their crystallographic position. It may be noted that, while the Mn-Ni distance obtained from Mn EXAFS remains unchanged, the same bond distance obtained from Ni EXAFS decreases from 2.58\AA~ to 2.56\AA~ with the decrease in temperature. Furthermore, the Ni-In bond increases by almost 0.07\AA~ in comparison to its room temperature value. This implies that within the tetrahedral arrangement around Ni atoms, the large movement of Mn atoms causes the Ni-Mn bonds to shrink at the cost of Ni-In bonds leading to a stronger Ni-Mn hybridization. 

\section{Discussion}
The understanding of atomic re-arrangements that occur upon martensitic transformation and hence hybridization that results due to the phase change is vital in determining the physical properties of such ferromagnetic shape memory alloys. In the local structural study of Ni-Mn-Ga alloys it has been shown that at low temperature the Ni-Ga hybridization grows stronger in compari75.50.Cc; son to Ni-Mn causing tetrahedral distortions within the L1$_0$ sub-cell of the parent L2$_1$ structure \cite{pab}. Such  $p-d$ hybridization that develops between X and Z atoms of the X$_2$YZ Heusler structure in the low temperature phase leads to re-distribution of electrons causing the band Jahn-Teller effect. Whereas EXAFS measurements on Ni$_{50}$Mn$_{35}$Sn$_{15}$ presents unequal Ni-Mn and Ni-Sn bond lengths that results in a stronger hybridization between Ni and the Mn occupying the Z-sites \cite{pab2}. From the study of these two systems, it may be conjentured that in the Ni-Mn based Heusler alloys, it is the strong hybridization between the X and Z species that results in alteration of the band structure at Fermi level and leads to martensitic transformation. In Ni$_{50}$Mn$_{35}$In$_{15}$ the crystal structure being B2, there is a complete disorder between the Y and Z sites and the identification of Mn that hybridizes with Ni, as Mn at Y-site or Mn at Z-site is not possible. However, the fact that there exists some Mn-$d$ and Ni-$d$ hybridization is clear from the shorter Ni-Mn and longer Ni-In bonds. It is this $d-d$ hybridization that is responsible for martensitic transformation in Ni$_{50}$Mn$_{35}$In$_{15}$.   

A martensitic transformation occurs when the Fermi surface touches the Brillouin zone boundary \cite{web}. This implies that change in factors like chemical pressure (as a result of difference in atomic sizes) and the e75.50.Cc; /a value can cause the alteration of the Fermi surface driving such systems towards structural instabilities. A linear dependence of the T$_M$ on changing e/a has indeed been observed for the Ni-Mn-Ga alloys \cite{lans,vasil}. However, in the present case, Ni$_{50}$Mn$_{35}$Sn$_{15}$ that has an e/a value of 8.05 undergoes a martensitic transition at much lower temperature as compared to Ni$_{50}$Mn$_{35}$In$_{15}$ with an e/a = 7.9. One of the main differences between these two systems is the crystal structure in the austenitic phase. Ni$_{50}$Mn$_{35}$Sn$_{15}$ has a cubic L2$_1$ structure while the Ni$_{50}$Mn$_{35}$In$_{15}$ crystallizes in a B2 structure. This could be perhaps due to the size difference between In and Mn in Ni$_{50}$Mn$_{35}$In$_{15}$ and Sn and Mn atoms in Ni$_{50}$Mn$_{35}$Sn$_{15}$. The size difference between In and Mn atoms being larger, results in a greater amount of disorder and therefore a more disordered B2 structure. Therefore, at least in case of Ni-Mn-In and Ni-Mn-Sn systems the martensitic transition temperatures depend upon the structural disorder rather than e/a ratio. This disorder also affects the hybridization in the martensitic phase. The difference between Ni-Mn and Ni-In bond lengths in Ni$_{50}$Mn$_{35}$In$_{15}$ is much larger as compared to that observed in the corresponding Ni-Mn and Ni-Sn bond lengths in Ni$_{50}$Mn$_{35}$Sn$_{15}$. In this Sn containing sample the Ni-Mn and Ni-Sn bond lengths obtained from EXAFS analysis are 2.57\AA~ and 2.61\AA~ respectively, giving a difference of 0.04\AA~ \cite{pab2}. Whereas as 75.50.Cc; can be seen from Table \ref{tab:in-xf} the difference between Ni-Mn and Ni-In bond lengths is 0.14\AA. This clearly indicates that the structural disorder in Ni$_{50}$Mn$_{35}$In$_{15}$ results in a stronger Ni($3d$)-Mn($3d$) hybridization in the martensitic phase. Such a  hybridization is reponsible for a rearrangement of the $d$ electrons within the hybrid band resulting in lifiting of degeneracy and lowering of the symmetry. This being stronger in the case of Ni$_{50}$Mn$_{35}$In$_{15}$ as compared to Ni$_{50}$Mn$_{35}$Sn$_{15}$ results in higher martensitic transition temperature.

\section{Conclusion}
The local structure of a new ferromagnetic shape memory alloy, Ni$_{50}$Mn$_{35}$In$_{15}$, was studied using the EXAFS technique. Comparision of the bond distance values between different atomic constituents in the austenitic and martensitic phase is made. A substantial off-center distortion of the tetrahedral arrangement around Ni atoms leading to a stonger Ni-Mn hybridization is observed in the martensitic phase. This hybirdization is more intense in Ni$_{50}$Mn$_{35}$In$_{15}$ than in similar composition, Ni$_{50}$Mn$_{35}$Sn$_{15}$ and in direct relation with the observed martensitic transformation temperatures in the two alloys.

\ack
Authors gratefully acknowledge financial assistance from Department of Science and Technology, New Delhi, India and ICTP-Elettra, Trieste, Italy for the proposal 2005743. Thanks are also due to Dr. Luca Olivi for help in EXAFS measurements. P.A.B. would like to thank Council for Scientific and Industrial Research, New Delhi for financial assistance.

\newpage

\begin{table}
\caption{\label{tab:in-xf}Mn and Ni K-edge EXAFS analysis of Ni$_{50}$Mn$_{35}$In$_{15}$ in the cubic and martensitic phase. R$_{calc}$ refers to the bond distance calculated for B2 cell from the lattice parameters obtained from XRD profile. R is the bond length as obtained from EXAFS analysis and $\sigma^2$ is the corresponding thermal mean square variation in the bond length. The analysis were carried out in $k$ range: (2-12)\AA$^{-1}$ with $k$-weight: 3 and $R$ range: (1-5)\AA ~for Mn edge and (1-3)\AA~ for Ni edge. Figures in parentheses indicate uncertainity in the last digit.}
\lineup
\begin{indented}
\item[]\begin{tabular}{@{}lllllll}
\br
\multicolumn{4}{c}{Cubic B2 phase} & \multicolumn{3}{c}{Martensitic phase} \\
Atom and & & & & Atom and & & \\
Coord. No. & R$_{calc}$ (\AA) & R (\AA) & $\sigma^2$ (\AA$^2$) & Coord. No. & R (\AA) & $\sigma^2$ (\AA$^2$)\\
\mr 
\multicolumn{7}{c}{Mn K-edge} \\
Ni1 $\times$ 8 & 2.615 & 2.567(3) & 0.0115(3) & Ni1 $\times$ 8 & 2.568(3) & 0.0079(3) \\
In1 $\times$ 1.8 & 3.020 & 2.886(5) & 0.0058(5) & In1 $\times$ 1.8 & 2.879(6) & 0.0046(6) \\
Mn1 $\times$ 4.2 & 3.020 & 2.911(5) & 0.0108(6) & Mn $\times$ 4.2 & 2.896(6) & 0.0094(7) \\
Mn2 $\times$ 12 & 4.270 & 4.27(4) & 0.035(7) & Mn1 $\times$ 8 & 4.19(1) & 0.016(2) \\
MS$^a \times$ 4.8 & 5.230 & 5.10(3) & 0.017(4) & Mn2 $\times$ 4 & 4.40(2) & 0.012(2)\\
MS$^b \times$ 11.2 & 5.230 & 5.175(6) & 0.0129(7) & Ni2 $\times$ 8 & 4.736(7) & 0.009(8)\\
& & & & Ni3 $\times$ 16 & 4.918(8) & 0.0080(9)\\
& & & & & & \\
\multicolumn{7}{c}{Ni K-edge} \\
Mn1 $\times$ 5.6 & 2.615 & 2.580(4) & 0.0128(5) & Mn1 $\times$ 5.6 & 2.558(3) & 0.0068(4)\\
In1 $\times$ 2.4 & 2.615 & 2.634(4) & 0.0070(4) & In1 $\times$ 2.4 & 2.70(1) & 0.009(1)\\
Ni1 $\times$ 6 & 3.020 & 3.13(6) & 0.038(9) & Ni1 $\times$ 2 & 2.797(7) & 0.0038(8)\\
& & & & Ni2 $\times$ 4 & 3.13(8) & 0.027(13) \\
\br
\end{tabular}
\item[]$^a$Mn$\rightarrow$In3$\rightarrow$Ni1$\rightarrow$Mn
\item[]$^b$Mn$\rightarrow$Mn3$\rightarrow$Ni1$\rightarrow$Mn
\end{indented}
\end{table}

\end{document}